# Observation of a Smoothly Tunable Dirac Point in Ge(Bi$_x$Sb$_{1-x}$)$_2$Te$_4$

Sean Howard[1], Arjun Raghavan[1], Davide Iaia[1], Caizhi Xu[1], David Flötotto[1], Man-Hong Wong[1], Sung-Kwan Mo[2], Bahadur Singh[3], Raman Sankar[4], Hsin Lin[4], Tai-Chang Chiang[1], and Vidya Madhavan[1*]

1. Department of Physics and Frederick Seitz Materials Research Laboratory, University of Illinois at Urbana-Champaign, Urbana, IL, USA
2. Advanced Light Source, Lawrence Berkeley National Laboratory, Berkeley, CA, USA
3. Department of Condensed Matter Physics and Materials Science, Tata Institute of Fundamental Research, Mumbai 400005, India
4. Institute of Physics, Academia Sinica, Taipei, Taiwan 11529

*vm1@illinois.edu

## Abstract

State-of-the-art topological devices require the use topological surface states to drive electronic transport. In this study, we examine a tunable topological system, Ge(Bi$_x$Sb$_{1-x}$)$_2$Te$_4$, for a range of *x* values from 0 to 1, using a combination of Fourier Transform Scanning Tunneling Spectroscopy (FT-STS) and Angle-Resolved Photoemission Spectroscopy (ARPES). Our results show that the Dirac point shifts linearly with *x*, crossing the Fermi energy near *x* = 0.7. This novel observation of a smoothly tunable, isolated Dirac point crossing through the topological transport regime and having strong linear dependence with substitution can be critical for future topological spintronics applications.

## Introduction

Since the discovery of topological materials, much effort has been made to utilize the properties of topologically protected states in novel electronic devices[1-7]. Among the primary challenges in realizing such devices is the isolation of topological surface states from a background of trivial bulk states[1-3]. If the chemical potential intersects bulk valence or conduction bands, transport will be dominated by bulk contributions[8]. To avoid this, the chemical potential must be manipulated into the band gap containing topological states, while keeping the band topology non-trivial.

One method to achieve transport dominated by topological states is chemical substitution. This type of topological engineering has been previously studied in material systems including $Bi_{2-x}Sb_xTe_{3-y}Se_y$, $(Bi_{1-x}Sb_x)_2Te_3$, and $BiTl(S_{1-x}Se_x)_2$[9-11]. In these materials, substitution moves the system into the topological transport regime, where transport is dominated by spin polarized topological states[9,11]. However, the systems previously studied have several challenges[12] associated with them, including difficulty in growth of materials and the lack of a smoothly tunable, isolated Dirac point. This motivates the study of new materials for topological engineering, and in this work, we study one such system, $Ge(Bi_xSb_{1-x})_2Te_4$.

Several properties of $Ge(Bi_xSb_{1-x})_2Te_4$ suggest the potential for tuning its topological states by substitution, making it ideal for optimizing the band structure to achieve transport dominated by the topological bands. Of the two parent compounds, $GeBi_2Te_4$ and $GeSb_2Te_4$, of this system, $GeBi_2Te_4$ is known to be a topological insulator with a band gap of ≈ 200 meV and an isolated Dirac point 280 meV below the Fermi energy[13]. This is advantageous since an isolated Dirac point is a critical component of electrical transport dominated by topological surface states (TSS). On the other end, time- and spin-resolved ARPES have also recently shown $GeSb_2Te_4$ to be topologically non-trivial, with a Dirac point approximately 450 meV above the Fermi energy[14].

Other material properties also contribute to technological interest in the system. Both parent compounds of this system are known thermoelectric materials[15]. In addition, $GeSb_2Te_4$ is a phase-change material which enters a metastable rock-salt phase by rapid cooling, suggesting possible use in nonvolatile random-access memory devices[16,17]. Finally, a large spin polarization is measured in the TSS of $GeBi_2Te_4$ by spin-resolved ARPES[18], which can be important for increased efficiency of spintronics devices. Detailed experimental studies of the effects of substitution in $Ge(Bi_xSb_{1-x})_2Te_4$ are therefore critically needed to understand its potential utility in a variety of electronic and spintronic applications.

## Experimental Methods

In this work, a combination of Fourier Transform Scanning Tunneling Microscopy (FT-STS) and Angle-Resolved Photoemission Spectroscopy (ARPES) is used to investigate $Ge(Bi_xSb_{1-x})_2Te_4$. ARPES measurements were carried out at the BL 10.0.1 at the Advanced Light Source in Lawrence Berkeley National Laboratory. To investigate the band structure of unoccupied states above the Fermi energy, and to confirm ARPES results below the Fermi energy, FT-STS is used. FT-STS measurements are performed at 4 K with a scanning tunneling microscope (STM) in ultrahigh vacuum; bulk single crystals are cleaved *in situ* and measured with annealed W STM tips. Details of crystal growth are given in the Appendix. Supplementary electrical transport measurements were performed using a 4-point contact method with a Quantum Design DynaCool Physical Property Measurement System.

The FT-STS technique utilizes quasiparticle interference (QPI) in the material, occurring due to quasiparticle scattering from defects and other imperfections, and yielding energy-dependent standing waves[19]. While direct backscattering is in general prohibited for topological surface states, due to the effects of hexagonal warping, a significant QPI signal is generated in $Ge(Bi_xSb_{1-x})_2Te_4$[12,20,21]. Similar warping effects have been previously observed in $Bi_2Te_3$[21] and $GeBi_2Te_4$[22], for example. Using the measured QPI scattering vectors, momentum-space band structure information is obtained. The combination of FT-STS and ARPES thus provides a powerful means to understand changes in the topological properties of the $Ge(Bi_xSb_{1-x})_2Te_4$ system.

**Results and Discussion**

$Ge(Bi_xSb_{1-x})_2Te_4$ crystals consist of stacked hexagonal layers forming a unit cell with c = 41.3 Å[22], as shown in Fig. 1(a). To characterize the surface, we obtain STM images of the cleaved crystals. Cleavage occurs between van der Waals bonded Te layers, as indicated in Fig. 1(a). A typical STM topography scan obtained on one of the materials in our series, $GeBi_{0.8}Sb_{1.2}Te_4$, shows the hexagonal lattice expected for these surfaces (Fig. 1(b)); topography scans for all other substitutions are shown in Appendix Fig. A3. Fig. 1(c) shows scanning tunneling spectroscopy (STS) dI/dV spectra, representing the local density of states (LDOS), for substitutions ranging from $x$ = 0.0 and $x$ = 1.0. The spectra are found to have major differences in locations of their minima. These reflect significant changes in the band structure due to substitution, and motivate further investigation using ARPES and FT-STS. More specifically, the Dirac point can be given[23] by the minimum of a dI/dV spectrum, which corresponds to the local density of states (LDOS). In this work, we show very strong agreement between Dirac point values found from our ARPES measurements, previously published ARPES spectra, and our STM dI/dV measurements. At $x$ = 1.0, the STM dI/dV minimum occurs at -280 meV, in agreement with the -280 meV ARPES measurement[22] by Arita, et al. (2014). For $x$ = 0.7, the lowest point of the LDOS is very close to the Fermi energy ($V_{Bias}$ = 0 mV), with a dI/dV minimum value of 4 meV. Further Sb substitution continues to monotonically shift the minimum to larger, positive energies, with $GeSb_2Te_4$ showing a dI/dV minimum at 492 meV. This is in agreement with the recent time- and spin-resolved ARPES results of Nurmamat, et al. (2020)[14], showing the Dirac point to be approximately 450 meV above the Fermi energy.

While the STM dI/dV results suggest that the $x$ = 0.7 substitution could place $Ge(Bi_xSb_{1-x})_2Te_4$ in the topological transport regime, a dI/dV minimum may not, in itself, be sufficient evidence. Hence, we next measure the band structure of Ge(Bi$_x$Sb$_{1-x}$)$_2$Te$_4$ for the compositions $x$ = 1, 0.7, 0.6, 0.5, and 0.4 using ARPES. Fig. 2(a) shows ARPES data for $x$ = 1, that is, GeBi$_2$Te$_4$, along the Γ-K direction. The intensity below a binding energy of 0.6 eV is from the bulk valence band. We also observe the contribution from the conduction band near the chemical potential. Surface states with approximate linear dispersion are seen to cross at a Dirac point near the zone center. The ARPES spectra for substitutions Ge(Bi$_x$Sb$_{1-x}$)$_2$Te$_4$ from $x$ = 0.7, down to $x$ = 0.4, are shown in Fig. 2(b)-(e). At all compositions, we observe the presence of quasi-linearly dispersing in-gap states, with $x$ = 1.0 and $x$ = 0.7 data allowing for linear fitting; this fitting is shown in Appendix Fig. A4. We confirm that the in-gap states seen are 2-dimensional surface states by varying the incoming photon energy between 62 eV and 70 eV. As demonstrated for GeBi$_{1.4}$Sb$_{0.6}$Te$_4$ in Fig. 2(f)-(h), with changing phonon energy, the in-gap states do not change significantly, while there are clear changes in the bulk valence bands at higher binding energies. To obtain further insights into the evolution of the band structure with Sb substitution, we shift the relative position of E$_F$ by depositing potassium on the surface (Fig. 2(i)-(j)). This allows us to see the lower part of the Dirac cone clearly, although further confirmation of the band structure above the Fermi energy is still necessary.

For visualizing the complete Dirac cone, and for comparison with ARPES data, FT-STS quasiparticle interference (QPI) scattering is employed. Slices of local density of states (LDOS) maps in GeBi$_{1.4}$Sb$_{0.6}$Te$_4$ at 200 meV and 100 meV, of size 60 nm × 60 nm, are shown in Fig. 3(a)-(b). We perform Fourier Transforms of each of these maps for analysis. In Ge(Bi$_x$Sb$_{1-x}$)$_2$Te$_4$, scattering between the hexagonally warped surface bands yields a significant QPI signal[20]. The states are identifiable as surface states from their close resemblance to related materials such as Bi$_2$Te$_3$, along with their symmetry[21,23]. Due to the spin texture, the signal is dominated by scattering across vectors connecting alternating Γ-K bands, resulting in scattering vectors along the Bragg peak Γ-M direction[20]; this scattering geometry is shown in Fig. 3(e). In Fig. 3(c)-(d), the Fourier transforms of the LDOS maps at 200 meV and 100 meV are presented. The Bragg peaks are visible at the corners of the FTs, and it is clear that the scattering wavevectors occur only in these Γ-M directions. Moreover, a significant change in the wavevectors is seen as a function of energy: as the energy is increased, the wavevectors get larger, indicating a positive dispersion. In the smaller energy 100 meV to -100 meV range, Fourier transforms of dI/dV slices do not show clear scattering vectors. However, the center of the slices, representing bulk band scattering, changes in size. Plotting a dispersion, as given in Fig. 3(f), of the slices in this energy range shows a clear gap of approximately 45 meV, crossing through 0 meV, consistent with our ARPES and dI/dV spectra data which suggest that the Dirac point for $x$ = 0.7 lies very close to the Fermi energy.

We can now compare data from the STM dI/dV, FT-STS, and ARPES measurements. The Dirac point energy is shown to have a strong linear dependence with doping; the Dirac point values and fitting are shown in Fig. 4(b). An $R^2$ value of 0.98 is obtained for the linear fit. As shown in the schematic of Fig. 4(a), and proven using a combination of FT-STS and ARPES, the topologically non-trivial band structure remains throughout the range of substitutions. The Dirac point crosses the Fermi energy very close to the $x = 0.7$ doping. This combination of an isolated Dirac point, along with its location within only 10 meV of the Fermi energy makes GeBi$_{1.4}$Sb$_{0.6}$Te$_4$ an ideal alloy for topological transport and spintronic[5] applications.

**Conclusion**

In conclusion, this detailed study employing a combination of STM, ARPES, and FT-STS has experimentally revealed a novel tunable topological insulator system, Ge(Bi$_x$Sb$_{1-x}$)$_2$Te$_4$, a valuable addition to existing work on other pseudobinary systems[24-28]. The ability to tune the smoothly tune the Dirac point as a linear function of substitution, without abrupt changes found in other related systems[13], will allow for new topological transport applications[29] for this series of materials. Future work, especially on the GeBi$_{1.4}$Sb$_{0.8}$Te$_4$ alloy, which resides at an ideal position in the topological transport regime, can examine technologically significant properties such as topological transmission through surface barriers, similar to previous measurements, for example, on Sb(111)[30]. The topology of Ge(Bi$_x$Sb$_{1-x}$)$_2$Te$_4$ thus provides a unique platform for a number of novel fundamental studies and technological applications.

**Appendix**

*Crystal Growth*

The highly oriented single crystals of Ge(Bi$_x$Sb$_{1-x}$)$_2$Te$_4$ are grown by using the vertical Bridgman method. Initially, stoichiometric mixtures of Ge(Bi$_x$Sb$_{1-x}$)$_2$Te$_4$ compounds were prepared with high purity metal precursors (5N) of bismuth (Bi), antimony (Sb), germanium (Ge), and tellurium (Te). These initial powders of Ge(Bi$_x$Sb$_{1-x}$)$_2$Te$_4$ were calcinated at 800 °C for 24 hrs. in a crucible furnace and then cooled down to room temperature slowly. Further finely ground compounds were transferred in carbon coated quartz tubes for highly oriented single crystal growth under an Ar atmosphere purged environment, and sealed under a vacuum of ≈ $2\times10^{-3}$ Torr. These double sealed quartz tube powders were then heated at 900 °C for 200 hrs. in a vertical Bridgman furnace to grow single crystals. The schematic of the crystal growth setup is shown in the inset of Fig. A1. A temperature gradient of ≈ 0.5 °C/mm is programmed at near the solidification point (marked by the red arrow in Fig. A1), and the quartz tubes are then moved slowly into the cooling zone with a translation rate close to 0.2 mm/hr. We have monitored and measured the temperature profile of the cooling process versus the relative position of the sample using an R-type thermocouple, as displayed in Fig. A1.

*Sample Stoichiometry*

Samples of different compositions are grown in this study. However, the nominal stoichiometry may differ from the actual stoichiometry[31]. To test that our actual compositions are close to the nominal values, we use an Electron Probe Micro-Analyzer (EPMA) technique.

| *x* - Nominal | *x* - Actual |
|---|---|
| 1.00 | 1.00 |
| 0.70 | 0.68 |
| 0.60 | 0.56 |
| 0.50 | 0.51 |
| 0.40 | 0.38 |
| 0.00 | 0.04 |

**Table A1 | Crystal Stoichiometries.** Compositions of Ge(Bi$_x$Sb$_{1-x}$)$_2$Te$_4$ samples are tested using an Electron Probe Micro-Analyzer (EPMA) technique. All measured compositions are very close to the nominal value with an average difference of only 4.4%.

As shown in Table A1 above, all crystal compositions are very close to the corresponding nominal values, yielding an average difference of only 4.4%. Hence, we confirm that our actual crystal compositions match well with nominal *x* values of 1.00, 0.70, 0.60, 0.50, 0.40, and 0.00.

*GeBi$_2$Te$_4$ Electrical Transport Measurements*

Electrical transport measurements are performed on GeBi$_2$Te$_4$ to further characterize the material, especially in the context of potential use in future technological applications. We find, as shown in Fig. A2, a relatively high mobility throughout the temperature range, with resistivity below 5 mΩ · cm, even at 300 K. The resistivity follows a monotonic decrease from 300 K down to 90 K, followed by an increase, and further decrease below approximately 50 K, consistent with previous resistivity data reported by Marcinkova, et al. (2013)[24].

*Ge(Bi$_x$Sb$_{1-x}$)$_2$Te$_4$ STM Topography Scans*

STM topography scans at all substitutions are presented in this paper; while *x* = 0.4 is shown in the main text, 20 nm x 20 nm scans for *x* = 0.0, 0.5, 0.6, 0.7, and 1.0 are given in Fig. A3. As is evident, there is a significant degree of electronic disorder across all substitutions. However, we were able to find large, flat atomically resolved areas for measurements at all substitutions.

*ARPES Spectra Fitting*

Linear fitting of ARPES spectra is shown in Fig. A4. We use MATLAB image processing and fitting capabilities to find high R$^2$ values above 0.9. Our linear fits yield Dirac point energy values of -272 meV for *x* = 1.0 and -6 meV for *x* = 0.7. The -272 meV value agrees well with previously reported ARPES results[22] on GeBi$_2$Te$_4$.

*STM dI/dV Spectra Minima*

We show, in Fig. A5, zoomed-in STM dI/dV spectra near the minima for each of the 6 substitutions measured. The full spectra are shown in Fig. 1(c). Our minima, which represent the energy of the Dirac point, are 104 meV, 230 meV, 268 meV, and 492 meV, for $x$ = 0.6, 0.5, 0.4, and 0.0. The $x$ = 0.0 value agrees with the ARPES-derived Dirac point value of ≈ 450 meV found by Nurmamat, et al. (2020)[14]. Moreover, for $x$ = 0.7, the STM dI/dV minimum is 4 meV, very close to the Fermi energy, and in agreement with our ARPES value of -6 meV. For $x$ = 1.0, the minimum is at -280 meV, again in very close agreement with our ARPES-measured value, and that of Arita, et al. (2014)[22].

## Acknowledgements

The STM work was supported by a grant from the U.S. Department of Energy (DOE), Office of Science (OS), Office of Basic Energy Sciences, under award No. DE-SC0014335. V.M. acknowledges partial support from the Gordon and Betty Moore Foundation's EPiQS Initiative through grant GBMF4860. This work was carried out, in part, at the Seitz Materials Research Laboratory Central Research Facilities, at the University of Illinois at Urbana-Champaign. The ARPES work is supported by the U.S. Department of Energy (DOE), Office of Science (OS), Office of Basic Energy Sciences, Division of Materials Science and Engineering, under Grant No. DE-FG02-07ER46383 (T-C.C.). This ARPES research used resources of the Advanced Light Source, which is a DOE Office of Science User Facility under contract No. DE-AC02-05CH11231. H.L. and R.S. acknowledge support from the Taiwan Ministry of Science and Technology (MOST), under Grant No. 109-2112-M-001-014-MY3 and Grant No. MOST-108-2112-M-001-049-MY2, respectively.

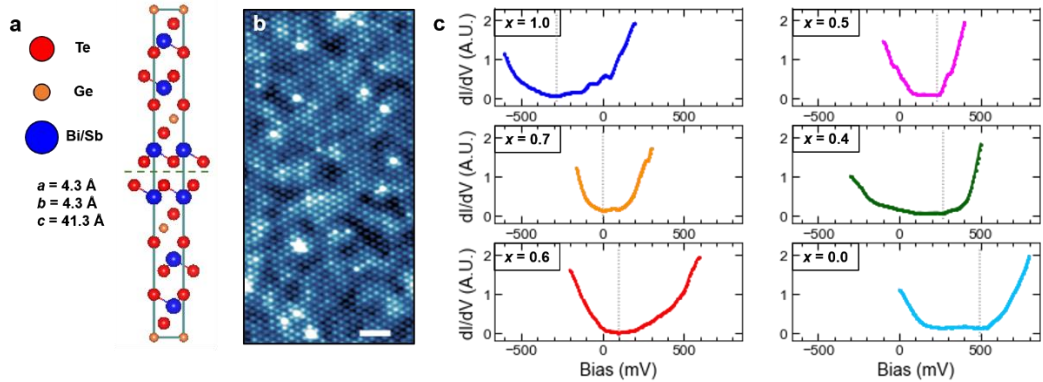

**Figure 1 | Crystal Structure, STM Topography, and dI/dV Spectra. (a)** Crystal structure of Ge(Bi$_x$Sb$_{1-x}$)$_2$Te$_4$; dashed line shows cleaving plane between two van der Waals bonded Te layers. **(b)** 10 nm × 20 nm topography ($I_T$ = 110 pA, $V_{Bias}$ = -400 mV) of Ge(Bi$_{0.4}$Sb$_{0.6}$)$_2$Te$_4$ showing a hexagonal lattice structure from the exposed Te surface. **(c)** STM dI/dV spectra for Bi substitutions of $x$ = 0.0, 0.4, 0.5, 0.6, 0.7, and 1.0; the differences between the spectra reflect changes in the band structure due to doping. It is clear that the minimum of the spectra shifts monotonically to higher energies with decreasing $x$, crossing the Fermi energy ($V_{Bias}$ = 0 mV) near $x$ = 0.7. The minima are marked by the light gray dotted lines.

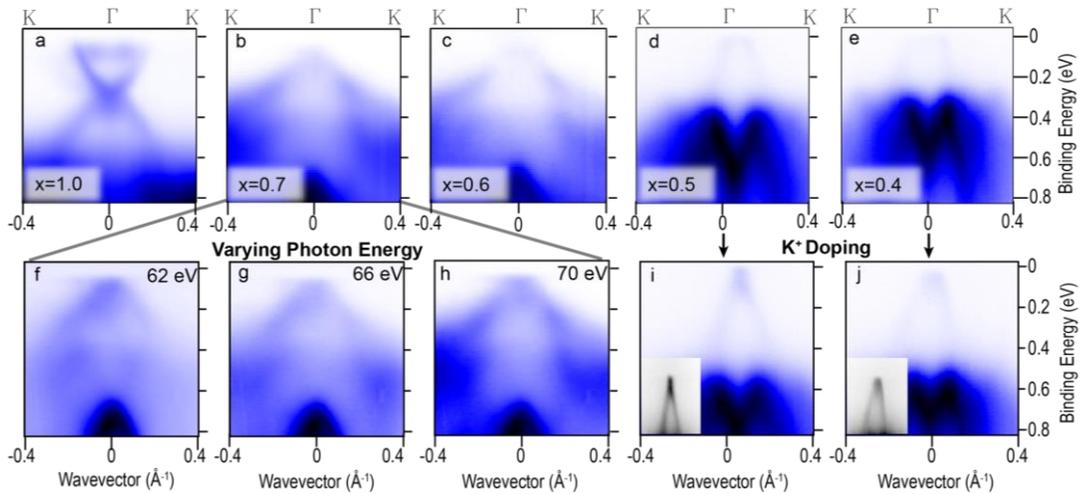

**Fig. 2 | ARPES Measurements. (a)-(e)** ARPES data along the Γ-K direction with different values of Bi doping $x$ for Ge(Bi$_x$Sb$_{1-x}$)$_2$Te$_4$. **(f)-(h)** GeBi$_{1.4}$Sb$_{0.6}$Te$_4$ ARPES spectra with varying incoming photon energy; the lack of change in the band structure suggests that the measured intensity is from 2D topological surface states. **(i)-(j)** ARPES measurements of potassium surface-doped samples with $x$ = 0.5 and $x$ = 0.4; K deposition shifts the Fermi level, making the Dirac point visible. Insets are close-ups of the spectra near the Dirac point.

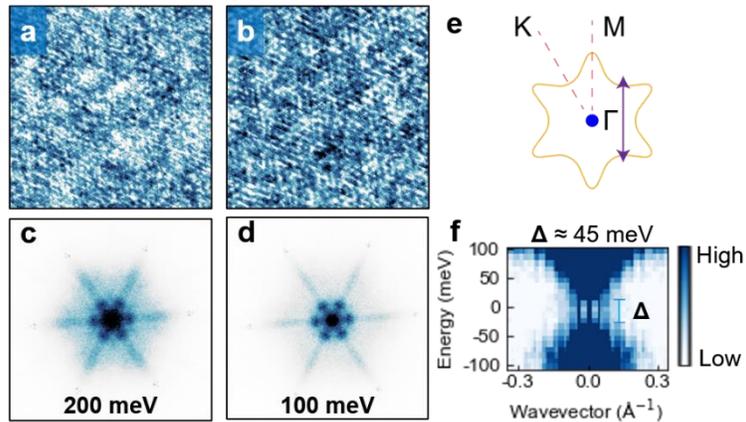

**Figure 3 | dI/dV Maps, Fourier Transforms, Scattering Schematic, and Bulk Band Dispersion in GeBi$_{1.4}$Sb$_{0.6}$Te$_4$. (a)-(b)** 60 nm$^2$ slices of a dI/dV map at 200 meV and 100 meV, respectively, for GeBi$_{1.4}$Sb$_{0.6}$Te$_4$. **(c)-(d)** Fourier transforms of dI/dV map slices in (a)-(b). The six Bragg peaks appear as faint dots near the outside of the slices, while the surface state quasiparticle interference (QPI) scattering vectors appear as brighter spots closer to the center. It is evident that the scattering vectors disperse with energy, while the Bragg peaks stay constant; the scattering vectors are identified as arising from the surface states due to their dispersion and close resemblance to the hexagonally-symmetric surface state QPI vectors in related materials such as Bi$_2$Te$_3$. **(e)** Schematic of hexagonally warped surface band structure; magenta arrow shows possible scattering vector along the Bragg peak Γ-M direction. **(f)** Bulk band dispersion from Fourier transforms of STM dI/dV map slices, found by plotting intensities along the Γ-M direction for each energy slice from a map in the +/- 100 meV range. We find a bulk gap of ≈ 45 meV traversing the Fermi energy.

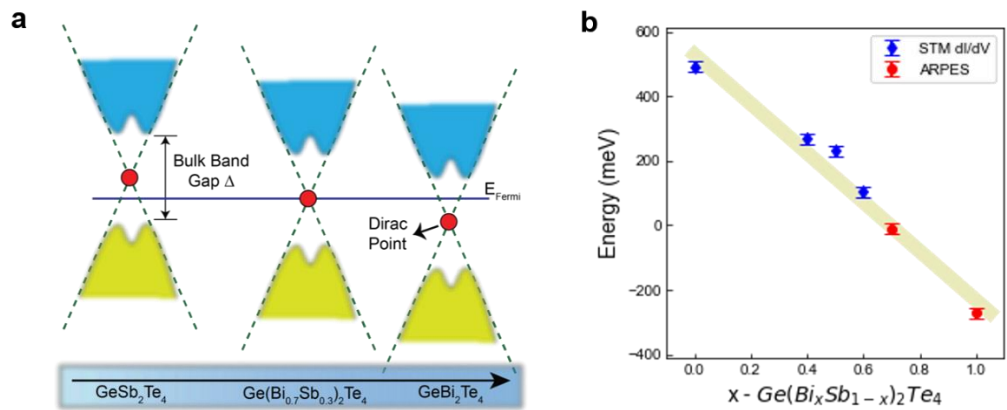

**Figure 4 | Band Structure Evolution and Shift of the Dirac Point. (a)** Schematic of band structure evolution as Sb is substituted with Bi. The linearly dispersing topological surface states intersect at the Dirac point, which shifts and crosses the Fermi energy near the $GeBi_{1.4}Sb_{0.6}Te_4$ substitution. **(b)** Plot of Dirac point energies, as determined by ARPES (red circles) and using the minima of STM dI/dV spectra (blue diamonds). The energies are shown to have a strong linear dependence on the substitution $x$ in $Ge(Bi_xSb_{1-x})_2Te_4$, with linear fitting yielding an $R^2$ value of 0.98.

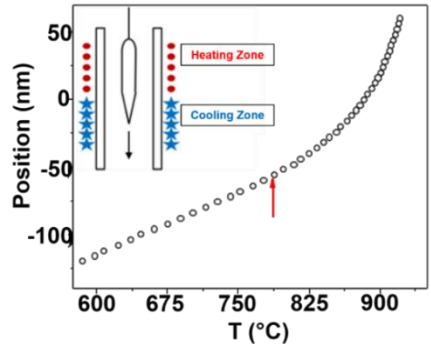

**Figure A1 | Single Crystal Growth.** Temperature profile of the cooling process versus the relative position of sample. Inset shows a schematic of our vertical Bridgman furnace.

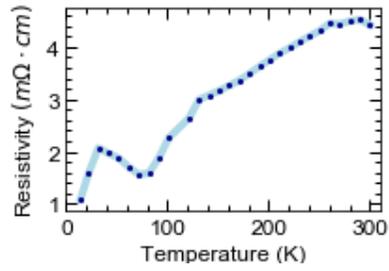

**Figure A2 | Electrical Transport Characterization of GeBi$_2$Te$_4$.** Electrical resistivity characterization of GeBi$_2$Te$_4$ shows relatively high mobility and an increase in resistivity as the temperature is decreased below 90 K, in close agreement with previous results from Marcinkova, et al. (2013)[24].

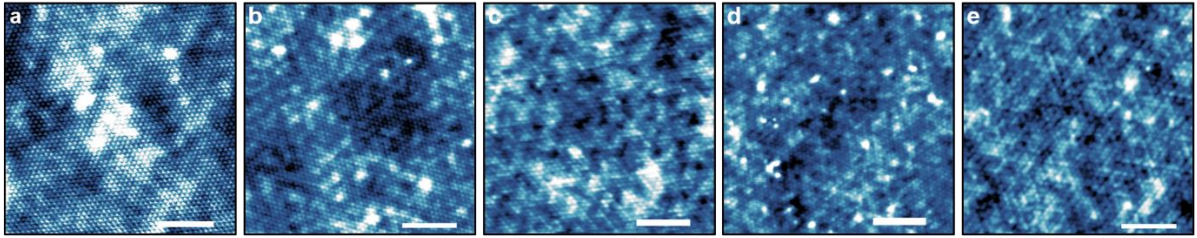

**Figure A3 | 20 nm × 20 nm Topography Scans at Different Substitutions. (a)** Topography of $GeSb_2Te_4$ with 80 pA and -100 mV. **(b)** Topography of $GeBiSbTe_4$ with 60 pA and -100 mV. **(c)** Topography of $GeBi_{1.2}Sb_{0.8}Te_4$ with 100 pA and 300 mV. **(d)** Topography of $GeBi_{1.4}Sb_{0.6}Te_4$ with 60 pA and -210 mV. **(e)** Topography of $GeBi_2Te_4$ with 360 pA and -250 mV.

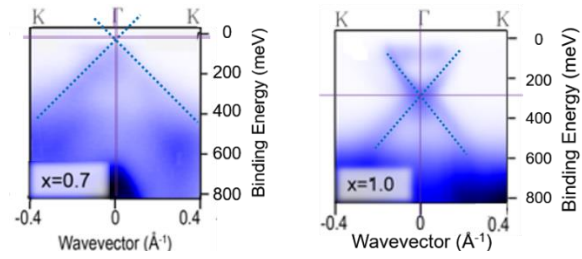

**Figure A4 | ARPES Fitting.** Linear fitting of ARPES spectra, by using MATLAB for processing, yields high $R^2$ values above 0.90, and Dirac points of -6 meV for $x$ = 0.7, and -272 meV for $x$ = 1.0. For better fitting clarity, we use the 66 eV photon energy spectrum for the $x$ = 0.7 case. Fitting is only used for the two substitutions shown above, since the Dirac cone is not clear for other values. Hence, for $x$ = 0.6, 0.5, 0.4, and 0.0, the Dirac point energies are extracted from STM dI/dV spectra.

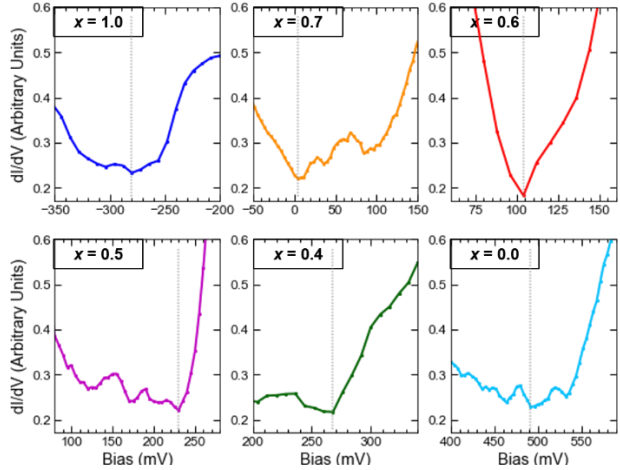

**Figure A5 | STM dI/dV Spectra Minima.** Zoomed-in STM dI/dV spectra close to the minima, for the six spectra presented in Fig. 1(c); gray dotted lines show the minimum energy values. While ARPES spectra are used to extract the Dirac point energy for $x$ = 1.0 and 0.7, the minima from the STM dI/dV spectra are used for $x$ = 0.6, 0.5, 0.4, and 0.0, yielding Dirac point energy values of 104 meV, 230 meV, 268 meV, and 492 meV, respectively. In addition, we find a minimum value of 4 meV for $x$ = 0.7, in close agreement with the Dirac point energy of -6 meV extracted from linear fitting of ARPES spectra.